\documentclass{umi}
\usepackage{amssymb}
\usepackage{amsmath}

\input{tcilatex}
\begin{document}

{\bf \centerline {The Hamilton principle for   fluid binary mixtures
with two temperatures}}

\bigskip

{\centerline  {Henri Gouin\footnote{University of Aix-Marseille \&
M2P2, C.N.R.S. U.M.R.  6181, Avenue Escadrille Normandie-Niemen,
  Box 322, 13397 Marseille Cedex 20  France.\ E-mail:  {henri.gouin@univ-cezanne.fr}}    and
    Tommaso Ruggeri \footnote{
Department of Mathematics and Research Center of Applied
Mathematics, C.I.R.A.M., University of Bologna, Via Saragozza 8,
40123-I Bologna Italy. E-mail: {ruggeri@ciram.unibo.it}; url:
{http://www.ciram.unibo.it/ruggeri}} }}
\bigskip

\bigskip

{\centerline {In "Bollettino della Unione Matematica Italiana" vol. 9 II (2009)}}

\bigskip
\bigskip

\begin{abstract}
\rm{F}or binary mixtures of fluids without chemical reactions, but
with
  components having different temperatures, the Hamilton principle
of least action is able to produce the equation of motion for each
component and a balance equation of  the total heat exchange between
components. In this nonconservative case, a  Gibbs dynamical
identity connecting the equations of momenta, masses, energy and
heat exchange
 allows to deduce the balance equation of energy of the mixture.
 Due to the unknown exchange of
heat between components, the number of obtained equations is less
than the number of field variables. The second law of
 thermodynamics constrains the possible expression of a supplementary
 constitutive equation   closing the system of
equations. The exchange of energy between components produces an
increasing rate of entropy and  creates a dynamical pressure term
associated with the difference of temperature between components.
This  new dynamical pressure term fits with the results obtained by
classical thermodynamical arguments in  \cite{GR} and confirms that
the Hamilton principle can afford to  obtain the equations of
motions for multi-temperature mixtures of fluids.

\end{abstract}


\section{Introduction.}

The theory of mixtures considers generally two different kinds of
continua: homogeneous mixtures (each component occupies the whole
mixture volume) and heterogeneous ones (each component occupies only
a part of the mixture volume). At least four approaches to the
construction of two-fluids models are known.\newline The first one
for studying the heterogeneous two-flows is an averaging method
(Ishii \cite{Ishii}; Nigmatulin \cite{Nigmatulin}). The averaged
equations of motion are obtained by applying an appropriate
averaging operator to the balance laws of mass, energy, etc...,
valid inside
each phase \cite{Drew,Win}. A second approach known as Landau method \cite%
{Bowen,Landau} was used for the construction of a quantum liquid
model and was purposed for the homogeneous mixtures of fluids
\cite{Putterman}. For the total mixture, the method requires the
balances of mass, momentum, energy, complemented with the Galilean
invariance principle and the second law of thermodynamics
\cite{Khalatnikov}. A third approach is presented in extended
thermodynamics; the mixtures are considered as a collection of
different media co-existing in the physical space. This approach is
done in the context of rational thermodynamics \cite{Truesdell}
founded on the postulate
that each constituent obeys the same balance laws as a single fluid \cite%
{Muller,ET,RS}. The thermodynamical processes must verify the second
law of thermodynamics and it is possible to purpose phenomenological
constitutive equations which allow to obtain the structure of
constitutive or production terms (as Fick's law, Fourier's Law etc.)
and to close the system of equations.

There exists a different approach based on the Hamilton principle
which is used for the construction of conservative (non-dissipative)
mathematical models
of continua. The principle was initiated by Lin \cite{Lin}, Serrin \cite%
{Serrin} and many others to obtain the governing equations of one component
continua \cite{Serrin,gouin0,gouin1} and involves an Hamilton action. The
variations of the Hamilton action are constructed in terms of virtual
motions of continua which may be defined both in Lagrangian and Eulerian
coordinates \cite{Serrin,gouin0}.\newline
Here, we use variations in the case of fluid mixtures. The variational
approach to the construction of two-fluid models has been used by many
authors (Bedford \& Drumheller \cite{Bedford}; Berdichevsky \cite%
{Berdichevsky}; Geurst \cite{Geurst2}; Gouin \cite{gouin2};
Gavrilyuk \& Gouin  \cite{Gavrilyuk 1}; Gouin \& Ruggeri
\cite{Gouin-Ruggeri}).
\newline To study thermodynamical processes by the Hamilton principle,
the entropy of the total mixture or the entropies of components are
added to the field parameters instead of temperatures. The
Lagrangian is the difference between the kinetic energy and an
internal potential per unit volume depending on the densities, the
entropies and the relative velocities of the mixture components (and
a potential due to external forces). The internal potential per unit
volume can be interpreted as a Legendre transformation of the
internal energy. In this case, it is not necessary to distinguish
 molecular mixtures from heterogeneous fluids when each component
occupies only a part of the mixture volume \cite{gouin3,Gavrilyuk
2}. Consequently, the terms including interaction between different
components of the mixture do not require constitutive postulates
difficult to interpret experimentally. They come from the direct
knowledge of the internal potential per unit volume. \newline The
assumption of a common temperature for all the components is open to
doubt for the suspensions of particles \cite{lhuillier} as well as
in the mixtures of gases in the early universe \cite{Groot}. By
using the Hamilton principle, the existence of several temperatures
(one temperature for each component) must be associated with the
existence of several entropies (one specific entropy for each
component). That will be the aim of this paper: the internal
potential per unit volume is a function of the densities, the
entropies and the difference of velocity between components.\newline
The plan of the article is as follows: \newline In section 2, we
formulate an extended form of the Hamilton principle of stationary
action allowing to produce the governing equations of motion for
each component of a binary mixture. From the invariance of time, we
deduce an equation of the exchange of heat between components. With
two temperatures (one temperature for each component), the system of
governing equations together with the balance of masses is not
closed: Hamilton's principle is not able to get a complete set of
equations when exchange of energies occurs between components. In
this case, we need an additive constitutive equation to close the
system.
\newline In section 3, we obtain a Gibbs dynamical identity. This
identity allows to obtain  the equation of energy for the total
mixture. The equation of motion for the total mixture and the
equation of energy are in divergence form; as in \cite{Stewart}, it
is not the same for the component equations of motion. \newline In
section 4, we consider the case of weakly dissipative mixtures and
introduce an average temperature of the total mixture. The average
temperature corresponds to a local equilibrium different from the
real state, with nonequal component temperatures but with the same
total internal energy. The total pressure of the mixture in the real
state is different from
the pressure associated with the mixture at the local equilibrium \cite%
{Ruggeri3,Ruggeri2}. The entropy variation rate of the mixture must
be in accordance with the second law of thermodynamics which implies
an additive constitutive equation for the pressure. The constitutive
equation depends on the physical properties of the two constituents
and we focus on the fact that the new pressure term does not obey
the same rule than the other terms due to dissipation.
\newline
 In section 5, as an
example, we reconsider the special case of   a mixture of perfect
 gases.\newline
 Finally, we compare our results   with the conclusions obtained
in \cite{GR} by using classical arguments of rational
thermodynamics.
\section{Governing equations in conservative cases.}

In a Galilean system of coordinates, the motion of a two-fluid continuum can
be represented by two diffeomorphisms
\begin{equation*}
\mathbf{Z}_{\alpha }=\mathbf{\Phi }_{\alpha }(\mathbf{z})\,,\qquad\quad
(\alpha=1, 2)  \label{representation}
\end{equation*}
or
\begin{equation*}
\lambda =t\ \ \ \ \ \mathrm{{and}\ \ \ \ \ }\mathbf{X}_{\alpha }=\mathbf{%
\phi }_{\alpha }(t,\mathbf{x)},  \label{representation2}
\end{equation*}
where $\mathbf{z}=(t,\mathbf{x)}$ denotes Eulerian coordinates in a
four-dimensional domain $\omega$ in the time-space and $\mathbf{Z}%
_\alpha=(\lambda,\mathbf{X}_\alpha)$ denotes Lagrangian coordinates of the
component $\alpha$ in a four-dimensional reference space $\omega_\alpha$.
The conservation of matter for each component requires that
\begin{equation}
\rho _{\alpha }\text{ det}\,F_a =\rho _{\alpha 0}\left( \mathbf{X}_{\alpha
}\right)\quad \mathrm{with}\quad F_a= \frac{\partial \mathbf{x}}{%
\partial \mathbf{X}_{\alpha }},  \label{mass0}
\end{equation}
where $\rho _{\alpha 0} $ is the reference density in $\omega_{\alpha }$ and
$\text{ det}\left( {\partial \mathbf{x}}/{\partial \mathbf{X}_{\alpha }}%
\right) $ is the Jacobian determinant of the motion of the component $\alpha$
of density $\rho _{\alpha }$. In differentiable cases equations (\ref{mass0}%
) are equivalent to the equations of density balances
\begin{equation}
{\frac{\partial \rho _{\alpha }}{\partial t}}\ +\ \text{div}(\rho _{\alpha }{%
\mathbf{v}}_{\alpha })=0 ,  \label{mass i}
\end{equation}
where ${\mathbf{v}}_{\alpha }$ denotes the velocity of each component $%
\alpha $.

The Lagrangian of the binary system is
\begin{equation*}
L = \sum_{\alpha =1}^{2}\, \left(\frac{1}{2} \ \rho _{\alpha }\,{\mathbf{v}}%
_{\alpha }^{2}-\rho _{\alpha }\Omega _{\alpha }\right) -\eta (\rho _{1},\rho
_{2},s_{1},s_{2},{\mathbf{u}}),
\end{equation*}
where the summation is taken over the fluid components $(\alpha =1,\,2)$ and
$\,s_{\alpha }$ are the specific entropies, $\,{\mathbf{u}}\,=\,{\mathbf{v}}%
_{2}-{\mathbf{v}}_{1}\ $ is the relative velocity of components, $\,\Omega
_{\alpha }\ $ are the external force potentials, $\,\eta $ is a potential
per unit volume of the mixture. The Lagrangian $\,L\,$ is a function of $%
\rho _{\alpha },{\mathbf{v}}_{\alpha },s_{\alpha }$ and we introduce the
quantities
\begin{eqnarray}
&&R_{\alpha }\ \equiv \ \displaystyle\frac{\partial L}{\partial \rho
_{\alpha }}\ =\frac{1}{2}\ {\mathbf{v}}_{\alpha }^{2}\,-\frac{\partial \eta
}{\partial \rho _{\alpha }}\ -\ \Omega _{\alpha },  \label{R} \\
&&{\mathbf{k}}_{\alpha }^{T}\ \equiv \ \displaystyle\frac{1}{\rho _{\alpha }}%
\ \frac{\partial L}{\partial {\mathbf{v}}_{\alpha }}\ =\ {\mathbf{v}}%
_{\alpha }^{T}\ -\frac{(-1)^{\alpha }}{\rho _{\alpha }}\ \frac{\partial \eta
}{\partial {\mathbf{u}}},  \label{k} \\
&&\rho _{\alpha }\ T_{\alpha }\ \equiv \ \displaystyle-\frac{\partial L}{%
\partial s_{\alpha }}\ =\frac{\partial \eta }{\partial s_{\alpha }},
\label{T}
\end{eqnarray}
where $^T$ denotes the transposition and $\displaystyle
\frac{\partial L}{\partial {\mathbf{v}}_{\alpha }}\,,\ \frac{\partial \eta }{%
\partial {\mathbf{u}}}$ are linear forms. Equation (\ref{T}) defines the
temperatures $T_{\alpha } \ (\alpha=1,2)$ which are dynamical quantities
depending on $\rho_1,\rho_2,s_1,s_2$ and ${\mathbf{u}}$.\newline
To obtain the equations of component motions by means of the Hamilton
principle, we consider variations of particle motions in the form of
surjective mappings,
\begin{equation*}
\mathbf{X}_{\alpha }=\mathbf{\Xi }_{\alpha }(t,\mathbf{x;}\varkappa _{\alpha
}),
\end{equation*}%
where scalars $\varkappa _{\alpha }$ are defined in a neighborhood of zero;
they are associated with a two-parameter family of virtual motions. The real
motions correspond to $\varkappa _{\alpha }=0\, $ such that $\mathbf{\Xi }%
_{\alpha }(t,\mathbf{x;}0)=\phi _{\alpha }(t,\mathbf{x)}$; the associated
virtual displacements generalize what is obtained for a single fluid \cite%
{gouin0,gouin3},
\begin{equation}
{\delta _{\alpha }}\mathbf{X}_{\alpha }=\frac{\partial \mathbf{\Xi }_{\alpha
}(t,\mathbf{x;}\varkappa _{\alpha })}{\partial \varkappa _{\alpha }}%
\left\vert _{\varkappa _{\alpha }=0}\right. .  \label{displacement1}
\end{equation}%
The Hamilton action is
\begin{equation*}
a=\int_{\omega }\ L\ dv\,dt.
\end{equation*}%
\emph{We first consider the Hamilton principle in the form}
\begin{equation*}
{\delta _{\alpha }}a\equiv \left( \frac{da}{d\varkappa _{\alpha }}\right)
_{|_{\varkappa _{\alpha }=0}}=\ {\delta _{\alpha }}\int_{\omega }\ L\
dv\,dt\ =\ 0,
\end{equation*}%
under constraints (\ref{mass0}), with ${\delta _{\alpha }}a\,$ being the
variations of $a$ associated with equation (\ref{displacement1}). \newline
From the definition of virtual motions, we obtain in Appendix A the values
of $\ {\delta _{\alpha }}{\mathbf{v}}_{\alpha }\,({\mathbf{x}},t)$, ${\delta
_{\alpha }}\rho _{\alpha }\,({\mathbf{x}},t)$ and ${\delta _{\alpha }}%
s_{\alpha }\,({\mathbf{x}},t)\,\;$where ${\delta _{\alpha }}\upsilon(t,{%
\mathbf{x}}) $ is the variation of $\upsilon$ at $(t,\,{\mathbf{x}})$ fixed.
By taking into account the formulae in Appendix A and the definitions (\ref%
{R}-\ref{T}), we get
\begin{equation*}
{\delta _{\alpha }}a=\int_{\omega }\ \bigg(R_{\alpha }{\delta _{\alpha }}%
\rho _{\alpha }+\rho _{\alpha }\,{\mathbf{k}}_{\alpha }^{T}\,{\delta
_{\alpha }}{\mathbf{v}}_{\alpha }-\rho _{\alpha }T_{\alpha }{\delta _{\alpha
}}s_{\alpha }\bigg)dv\,dt
\end{equation*}%
\begin{equation*}
=\int_{\omega _{\alpha }}\bigg(R_{\alpha }\,\mathrm{div}_{\alpha }(\rho
_{\alpha 0}\,{\delta _{\alpha }}{\mathbf{X}}_{\alpha })-\rho _{\alpha 0}\,{%
\mathbf{k}}_{\alpha }^{T}\ F_{\alpha }\ \frac{\partial }{\partial \lambda }({%
\delta _{\alpha }}{\mathbf{X}}_{\alpha })-\rho _{\alpha 0}\,T_{\alpha }\
\frac{\partial s_{\alpha 0}}{\partial {\mathbf{X}}_{\alpha }}\ {\delta
_{\alpha }}{\mathbf{X}}_{\alpha }\bigg)dv_{\alpha }\,dt,
\end{equation*}%
where $\mathrm{div}_{\alpha }$ is the divergence operator with respect to
the coordinates ${\mathbf{X}_{\alpha }}$. In the last expression all
quantities are considered as functions of $\,(t,{\mathbf{X}}_{\alpha })$;
the functions are assumed to be smooth enough in the domain $\,\omega
_{\alpha }\,$ and $\ {\delta _{\alpha }}{\mathbf{X}}_{\alpha }\ =\ 0\ $ on
its boundary. Hence, we get
\begin{equation*}
{\delta _{\alpha }}a=\int_{\omega _{\alpha }} \rho _{\alpha 0}\left( -\frac{%
\partial R_{\alpha }}{\partial {\mathbf{X}}_{\alpha }}+\frac{\partial }{%
\partial \lambda }({\mathbf{k}}_{\alpha }^{T}\ F_{\alpha })-T_{\alpha }\frac{%
\partial s_{\alpha 0}}{\partial {\mathbf{X}}_{\alpha }}\right) {\delta
_{\alpha }}{\mathbf{X}}_{\alpha } \ dv_{\alpha }\,dt,
\end{equation*}%
and we obtain the equations of component motions in Lagrangian coordinates,
\begin{equation*}
\frac{\partial }{\partial \lambda }({\mathbf{k}}_{\alpha }^{T}\ F_{\alpha })-%
\frac{\partial R_{\alpha }}{\partial {\mathbf{X}}_{\alpha }}-T_{\alpha }\,%
\frac{\partial s_{\alpha 0}}{\partial {\mathbf{X}}_{\alpha }}\ =0,
\label{(equimage)}
\end{equation*}
where ${\mathbf{k}}_{\alpha }^{T}$ is defined by equation (\ref{k}).\newline
By taking into account the identity $\displaystyle\ {\frac{d_{\alpha
}F_{\alpha }}{dt}}\,-\,{\frac{\partial {\mathbf{v}}_{\alpha }}{\partial {%
\mathbf{x}}}}\ F_{\alpha }=0$ and for $\lambda =t$, we rewrite the equations
in Eulerian coordinates,
\begin{equation}
\frac{d_{\alpha }{\mathbf{k}}_{\alpha }^{T}}{dt}\ +\,{\mathbf{k}}_{\alpha
}^{T}\ \frac{\partial {\mathbf{v}}_{\alpha }}{\partial {\mathbf{x}}}\,=\,%
\frac{\partial R_{\alpha }}{\partial {\mathbf{x}}}\ +T_{\alpha }\frac{%
\partial s_{\alpha }}{\partial {\mathbf{x}}}\ .  \label{(motions)}
\end{equation}
The covector $\,{\mathbf{k}}_{\alpha }^{T}\,$ is an essential quantity;
indeed, $\rho _{\alpha }{\mathbf{k}}_{\alpha }\,$ (and not $\,\rho _{\alpha }%
{\mathbf{v}}_{\alpha }\,$) is the momentum for the component $\alpha $ of
the mixture.

To obtain the equation of energy, \emph{we need a second variation of
motions associated with the time parameter}. The variation corresponds to a
virtual motion in the form
\begin{equation*}
\lambda =\varphi (t;\kappa ),
\end{equation*}%
where scalar $\kappa $ is defined in a neighborhood of zero. The real motion
of the mixture corresponds to $\kappa =0$ such that $\varphi (t;0)=t$; the
associated virtual displacement is
\begin{equation*}
\delta \lambda =\frac{\partial \varphi (t;\kappa )}{\partial \kappa }%
\left\vert _{\kappa =0}\right. .
\end{equation*}%
For a single fluid, the entropy is defined on a reference space $\omega _{o}$
associated with Lagrangian variables; in conservative motion the specific
entropy is conserved along the trajectories and in the reference space the
entropy depends only on Lagrangian variables ${\mathbf{X}}$ and not on $%
\lambda $.\newline
In multi-component fluids, due to exchanges of energy between the
components, the entropies cannot be conserved along component paths; in the
reference spaces $\omega _{\alpha }$, the specific entropies $s_{\alpha }$
depend also on $\lambda $
\begin{equation*}
s_{\alpha }=s_{\alpha 0}\left( \lambda ,\mathbf{X}_{\alpha }\right) .
\end{equation*}%
\emph{The variation of Hamilton's action associated with the second family
of virtual motions} yields
\begin{equation*}
{\delta }a\equiv \ {\delta }\int_{\omega }\ L\ d\mathbf{x}\,dt\ =\
\int_{\omega }\ \frac{\partial L}{\partial \lambda }\ \delta \lambda \
dv\,dt\,=0 .
\end{equation*}%
From $\displaystyle  \frac{\partial L}{\partial \lambda }=\sum_{%
\alpha=1}^{2} \frac{\partial L}{\partial s_{\alpha }}\, \frac{\partial
s_{\alpha 0}}{\partial \lambda }$, we deduce when $\lambda=t$, $%
\displaystyle  \frac{\partial L}{\partial \lambda }= -\sum_{\alpha =1}^{2}\
\rho _{\alpha }T_{\alpha }\frac{d_{\alpha }s_{\alpha }}{dt}$, where\ $%
\displaystyle
\frac{d_{\alpha }s_{\alpha }}{dt}=\frac{\partial s_{\alpha }}{\partial t}+%
\frac{\partial s_{\alpha }}{\partial \mathbf{x}}\,\mathbf{v}_{\alpha }$ is
the material derivative with respect to velocity  $\mathbf{v}_{\alpha }$.
We obtain for the total mixture
\begin{equation}
\sum_{\alpha =1}^{2} \rho _{\alpha }T_{\alpha }\frac{d_{\alpha }s_{\alpha }}{%
dt} =0 .  \label{total energy01}
\end{equation}
Due to equations (\ref{mass i}) we obtain the equivalent form
\begin{equation}
\sum_{\alpha =1}^{2} Q_\alpha=0\qquad \mathrm{with} \qquad Q_\alpha=\left(
\frac{\partial \rho _{\alpha }s_{\alpha }}{\partial t}+\mathrm{{div}(\rho
_{\alpha }s_{\alpha }\mathbf{v}_{\alpha })}\right) \,T_{\alpha } .
\label{total energy1}
\end{equation}%
Equation (\ref{total energy1}) expresses that the exchange of energy between
components has a null total amount.

\section{Gibbs dynamical identity and equation of energy.}

Let us prove that equation (\ref{total energy1}) leads to the equation of
energy of the mixture. We introduce the quantities $\,{\mathbf{M}}_{\alpha
},\,B_{\alpha },S\,$ and $E$ such that
\begin{eqnarray*}
&&\displaystyle{\mathbf{M}}_{\alpha }^{T}\ =\ \rho _{\alpha }\frac{d_{\alpha
}{\mathbf{k}}_{\alpha }^{T}}{dt}\ +\ \rho _{\alpha }\ {\mathbf{k}}_{\alpha
}^{T}\frac{\partial {\mathbf{v}}_{\alpha }}{\partial {\mathbf{x}}}\ -\ \rho
_{\alpha }\frac{\partial R_{\alpha }}{\partial {\mathbf{x}}}\ -\rho _{\alpha
}\ T_{\alpha }\frac{\partial s_{\alpha }}{\partial {\mathbf{x}}}, \\
&&\displaystyle{B}_{\alpha }\ =\frac{\partial \rho _{\alpha }}{\partial t}\
+\ \mathrm{{div}(\rho _{\alpha }{\mathbf{v}}_{\alpha }),} \\
&&\displaystyle S\ =\sum_{\alpha =1}^{2} Q_\alpha, \\
&&\displaystyle E = \sum_{\alpha =1}^{2} \frac{\partial }{\partial t} \bigg(%
\rho _{\alpha }\left(\frac{1}{2}{\mathbf{v}}_{\alpha }^{2}+\Omega _{\alpha
}\right)+\eta -\frac{\partial \eta }{\partial {\mathbf{u}}}\,\mathbf{u}\bigg)%
+\mathrm{{div}\big(\rho _{\alpha }{\mathbf{v}}_{\alpha }({\mathbf{k}}%
_{\alpha }^{T}{\mathbf{v}}_{\alpha }-R_{\alpha })\big )-\rho _{\alpha }\frac{%
\partial \Omega _{\alpha }}{\partial t}}\,,
\end{eqnarray*}%
where $\displaystyle\eta -\frac{\partial \eta }{\partial {\mathbf{u}}}\,%
\mathbf{u} = f $ is the Legendre transformation of $\eta $ with respect to ${%
\mathbf{u}}$ and corresponds to the volume internal energy of the mixture.
We prove in Appendix B the following property:

\bigskip \noindent \textbf{Theorem:}\textit{\ \ For any motion of the
mixture, we have the algebraic identity}
\begin{equation}
E\ -\left( \,\sum_{\alpha =1}^{2} \ M_{\alpha }^{T}\ {\mathbf{v}}_{\alpha }+(%
{\mathbf{k}}_{\alpha }^{T}\ {\mathbf{v}}_{\alpha }-R_{\alpha }+T_{\alpha
}s_{\alpha })\ B_{\alpha }\right) -S\ \equiv 0 .  \label{Identity of Gibbs}
\end{equation}%
Relation (\ref{Identity of Gibbs}) is the general expression of the \textit{%
Gibbs identity} in dynamics. Analogous identities were obtained earlier for
thermocapillary mixtures \cite{gouin3} and bubbly liquids \cite{Gavrilyuk 3}%
. Due to the equations of balance of masses (\ref{mass i}), momenta (\ref%
{(motions)}) and energy (\ref{total energy1}), deduced from Hamilton's
principle, which are respectively
\begin{equation*}
B_{\alpha }=0,\quad {\mathbf{M}}_{\alpha }=0\ \ (\alpha =1,2)\quad \text{and}%
\quad S=0,  \label{conservative equations}
\end{equation*}%
we obtain from identity (\ref{Identity of Gibbs}):

\noindent \textbf{Corollary:} \emph{\ The motions of a mixture satisfy the
equation of energy balance in the form}
\begin{equation}
\sum_{\alpha =1}^{2} \frac{\partial }{\partial t} \bigg(\rho _{\alpha }
\left(\frac{1}{2}\,{\mathbf{v}}_{\alpha }^{2} +\Omega _{\alpha }\right)+ f %
\bigg) +\mathrm{{div}\bigg(\rho _{\alpha } {\mathbf{v}}_{\alpha }\, \left({%
\mathbf{k}}_{\alpha }^{\emph{T}}\, {\mathbf{v}}_{\alpha } - R_{\alpha
}\right) \bigg)-\rho _{\alpha } \frac{\partial \Omega _{\alpha }}{\partial t}%
} =0\, .  \label{total energy2}
\end{equation}%
Equation (\ref{total energy2}) appears as the equation of energy when $f$ is
the \emph{total internal energy}.

The equations of component motions are not written in divergence form.
Nevertheless by summing equations (\ref{(motions)}) in the form ${\mathbf{M}}%
_{\alpha }^{T}=0$ and taking into account equations (\ref{mass0}) in the
form $B_{\alpha } = 0$, we obtain by a calculation similar as in \cite%
{Gouin-Ruggeri,gouin3,Gavrilyuk 2} the \emph{balance equation for the total
momentum} in a divergence form:
\begin{equation}
\sum_{\alpha =1}^{2} \frac{\partial {\rho _{\alpha }\,\mathbf{v}}_{\alpha
}^{T}}{\partial t}\,+\,\mathrm{{div}\left(\rho _{\alpha }\,{\mathbf{v}}%
_{\alpha }\,{\mathbf{k}}_{\alpha }^{T}+\Big(\rho _{\alpha }\,\frac{\partial
\eta }{\partial \rho _{\alpha }}- \eta \Big)\ \mathbf{I}\right)\,+\rho
_{\alpha }\,\frac{\partial \Omega _{\alpha }}{\partial {\mathbf{x}}}} =0 ,
\label{totalmoment}
\end{equation}
where $\mathbf{I}$ is the identity tensor. In the following, $\displaystyle %
\rho\, {\mathbf{v}}= \sum_{\alpha =1}^{2}\rho _{\alpha }{\mathbf{k}}_{\alpha
}=\sum_{\alpha =1}^{2}\rho _{\alpha }{\mathbf{v}}_{\alpha }$ is the total
momentum and $\displaystyle\rho =\sum_{\alpha =1}^{2}\rho _{\alpha }\ $ is
the  mixture density.

System ((\ref{mass i}),(\ref{(motions)}),(\ref{total energy2})),
consequence of the Hamilton principle, is a non closed system of  equations.
In a single conservative fluid, the system of motion equations is closed by
the entropy conservation. In case of mixtures with two entropies, the
Hamilton principle is not able to close the system of motion equations; we
need additional arguments to obtain the evolution equations for each entropy
$s_{\alpha }$ by considering the behaviors of $Q_{\alpha }$.\newline
A possibility to close the system of equations is to consider the case when
the momenta and heat exchanges between the components are rapid enough to
have a common temperature. This case is connected with a conservative
equation for the total specific entropy \cite{gouin3}.\newline
Another possibility, used by Landau for quantum fluids \cite{Landau}, is to
assume that the total specific entropy $s$ is convected along the first
component trajectory
\begin{equation*}
{\frac{\partial \rho _{1}s}{\partial t}}\ +\ div\ (\rho _{1}s\, \mathbf{v}%
_{1})=0.
\end{equation*}%
In this case, the constitutive functions are $\rho _{1},s,\mathbf{v}%
_{1},\rho _{2},\mathbf{v}_{2}$, where $\rho _{\alpha }\ (\alpha=1,2)$ are
submitted to the constraints (\ref{mass i}) and the case of Helium
superfluid is a special case of our study corresponding to $s_{1}=s$ and $%
s_{2}=0$. Such an hypothesis is not acceptable for classical fluids. These
assumptions are not valid for heterogeneous mixtures where each phase has
different pressures and temperatures \cite{lhuillier,Groot}.\newline
In the following we consider the case when the mixture is \emph{weakly out
of equilibrium} such that the difference of velocities $\mathbf{u}$ and the
difference of temperatures $T_{2}-T_{1}$ are small enough with respect to
the main field variables.

\section{Mixtures weakly out of equilibrium.}

For the sake of simplicity, we neglect the external forces. Generally, the
volume potential $\eta $ is developed in the form \cite{Gavrilyuk
1,Gavrilyuk 3}\footnote{%
In \cite{GR}, the internal energy is the sum of the internal energies of the
components  $\left(\rho\,\varepsilon = \sum_{\alpha=1}^2
\rho_\alpha\varepsilon_\alpha(\rho_\alpha,s_\alpha)\right)$.}
\begin{equation*}
\eta (\rho _{1},\rho _{2},s_{1},s_{2},\mathbf{u})=e(\rho _{1},\rho
_{2},s_{1},s_{2})-b(\rho _{1},\rho _{2},s_{1},s_{2})\,\mathbf{u}^{2} ,
\end{equation*}%
where $b$ is a positive function of $\rho _{1},\rho _{2},s_{1},s_{2}$.
Properties of convexity of the function $\eta $ are studied in \cite%
{Gavrilyuk 2}. When $|\mathbf{u}|$ is small enough, the equations of motions
are hyperbolic \cite{Gavrilyuk 2}. We consider the linear approximation when
$|\mathbf{u}|$ is small with respect to $|\mathbf{v}_{1}|$ and $|\mathbf{v}%
_{2}|$. In linear approximation the volume potential is equal to the volume
internal energy $e$,
\begin{equation*}
\eta (\rho _{1},\rho _{2},s_{1},s_{2},\mathbf{u}) \approx e(\rho _{1},\rho
_{2},s_{1},s_{2})=\rho\,\varepsilon(\rho _{1},\rho _{2},s_{1},s_{2}),
\end{equation*}
where $\varepsilon$ denotes the internal energy per unit mass. Let us note
that the diffusion vector $\mathbf{j}=\rho _{1}(\mathbf{v}_{1}-\mathbf{v}%
)\equiv \rho _{2}(\mathbf{v}-\mathbf{v}_{2})$ is a small momentum vector
deduced respectively from velocities and densities of the components. The
equations of density balances can be written in the form
\begin{equation}
\frac{d\rho }{dt}+\rho \,\mathrm{div}\,\mathbf{v}=0 \qquad \mathrm{and}%
\qquad \rho \,{\frac{dc}{dt}}+\mathrm{div}\,{\mathbf{j}}=0 ,  \label{masses}
\end{equation}%
\newline
where $\displaystyle\, c=\frac{\rho _{1}}{\rho} $ denotes the concentration
of component $1$ and $\displaystyle \frac{\emph{d}\ }{\emph{dt}}=\frac{%
\partial \ }{\partial \emph{t}}+\frac{\partial \ }{\partial \mathbf{x}}\,.\,{%
\mathbf{v}}$ is the material derivative with respect to the average velocity
of the mixture. \newline
The divergence of a linear operator $\mathbf{A}$ is the covector $\mathrm{div%
} \mathbf{A }$ such that, for any constant vector ${\mathbf{a}}, $ $(\mathrm{%
div}\, \mathbf{A})\, {\mathbf{a}} = \mathrm{div }\, (\mathbf{A}\, {\mathbf{a}%
})$ and we write ${\mathbf{v}}_{\alpha }{\mathbf{v}}_{\alpha }^T \equiv{%
\mathbf{v}}_{\alpha }\otimes {\mathbf{v}}_{\alpha }$. \newline
Let us denote by $\displaystyle h_{\alpha }\equiv \frac{\partial e}{\partial
\rho _{\alpha }}\ $ the specific enthalpy of the component $\alpha$. \newline
For processes with weak diffusion, the equations of component motions get
the form,
\begin{equation*}
\rho _{\alpha }\,\mathbf{\Gamma }_{\alpha }\equiv \frac{\partial {\rho }_{{%
\alpha }}{\mathbf{v}}_{\alpha }}{\partial t}\,+\,\mathrm{div}(\rho _{\alpha }%
{\mathbf{v}}_{\alpha }\otimes {\mathbf{v}}_{\alpha })^{T}=\rho _{\alpha
}T_{\alpha }\,\mathrm{{grad}\,\emph{s}_{\alpha }-\rho _{\alpha }\,grad\,%
\emph{h}_{\alpha } }.  \label{component equation of motions i}
\end{equation*}%
The equation of total momentum (\ref{totalmoment}) is reduced to
\begin{equation*}
\frac{\partial {\rho \mathbf{v}}}{\partial t}\,+\,\mathrm{div}\left(
\sum_{\alpha =1}^{2}\left( \rho_\alpha {\mathbf{v}}_{\alpha }\otimes {%
\mathbf{v}}_{\alpha }\right) -\mathbf{t}\right) ^{T}=0,
\end{equation*}%
where $\ \mathbf{t}=\sum_{\alpha =1}^{2}\mathbf{t}_{\alpha }\ $ is the total
stress tensor such that
\begin{equation*}
\mathbf{t}_{\alpha \nu \gamma }=-p_{\alpha }\;\delta _{\nu \gamma },\;\text{%
with}\quad p_{\alpha }=\ \rho\rho_\alpha\varepsilon_{,\rho_\alpha}=\rho
_{\alpha }e_{,\rho _{\alpha }}-\dfrac{\rho _{\alpha }e}{\rho }\,,\quad
p=\sum_{\alpha =1}^{2}p_{\alpha }\, .  \label{pressure}
\end{equation*}
The equation of energy (\ref{total energy2}) writes in the simpler form
\begin{equation*}
\frac{\partial }{\partial t}\left( e+\sum_{\alpha =1}^{2}\frac{1}{2}\,\rho
_{\alpha }\mathbf{v}_{\alpha }^{2}\right) +\mathrm{div}\left( e\ \mathbf{v}%
+\sum_{\alpha =1}^{2}\left( \frac{1}{2}\,\rho _{\alpha }\mathbf{v}_{\alpha
}^{2}-\mathbf{t}_{\alpha }\right) \mathbf{v}_{\alpha }\right) =0.
\label{Total energy2}
\end{equation*}
The internal energy is a natural function of densities and entropies. Due to
equation (\ref{T}),
\begin{equation}
\rho_1\,T_{1}= \rho\, \frac{\partial \varepsilon}{\partial s_{1}}(\rho
_{1},\rho _{2},s_1,s_2)\quad \mathrm{and}\quad \rho_2\,T_{2}= \rho\, \frac{%
\partial \varepsilon}{\partial s_{2}}(\rho _{1},\rho _{2},s_1,s_2).
\label{Temperatures}
\end{equation}
Let us denote by $\overline{\varepsilon}$ the expression of the specific
internal energy as a function of $\rho, c, s_1, s_2$ such that $\overline{%
\varepsilon}(\rho,c,s_1,s_2)=  \varepsilon(\rho_1,\rho_2,s_1,s_2)$; we get
\begin{equation*}
\rho \,\frac{d\varepsilon }{dt}= \rho\, \frac{\partial \overline{\varepsilon}%
}{\partial \rho}\,\frac{d\rho }{dt}+\rho\, \frac{\partial \overline{%
\varepsilon}}{\partial c}\,\frac{dc }{dt}+\rho\, \frac{\partial \overline{%
\varepsilon}}{\partial s_1}\,\frac{ds_1}{dt}+\rho\, \frac{\partial \overline{%
\varepsilon}}{\partial s_2}\,\frac{ds_2}{dt}.  \label{Gibbsa}
\end{equation*}
Due to the fact that $\displaystyle \rho^2\, \frac{\partial \overline{%
\varepsilon}}{\partial \rho} = p\ $ and $\ \displaystyle
\frac{\partial \overline{\varepsilon}}{\partial c}= h_1-h_2$, we obtain
\begin{equation}
\rho \,\frac{d\varepsilon }{dt}= \frac{p}{\rho}\,\frac{d\rho }{dt}+\rho\,
(h_1-h_2)\,\frac{dc }{dt}+\rho_1\, T_1 \,\frac{ds_1}{dt}+\rho_2\, T_2\,%
\frac{ds_2}{dt}.  \label{Gibbsb}
\end{equation}
By taking into account that
\begin{equation*}
\frac{d_\alpha s_\alpha}{dt}=\frac{ds_\alpha}{dt}+ \frac{\partial s_\alpha}{%
\partial \mathbf{x}}(\mathbf{v}_\alpha-\mathbf{v})
\end{equation*}
and by using equations (\ref{total energy01}), (\ref{masses}), equation (\ref%
{Gibbsb}) yields
\begin{equation}
\rho \,\frac{d\varepsilon }{dt}+p\ \mathrm{div}\,\mathbf{v}\,+(h_{1}-h_{2})\
\mathrm{div}\,\mathbf{j}+(T_{1}\,\mathrm{{grad}\,\emph{s}_{1}-T_{2}\,{grad}\,%
\emph{s}_{2})^{\emph{T}}\,\mathbf{j}=0 }.  \label{Total energy4}
\end{equation}

Due to equations (\ref{Temperatures}), the internal energy can be expressed
as a function of densities and temperatures of components
\begin{equation*}
\tilde{\epsilon}(\rho _{1},\rho _{2},T_{1},T_{2})=\varepsilon (\rho
_{1},\rho _{2},s_{1},s_{2}).
\end{equation*}%
As we did in \cite{GR}, we define the average temperature $T$ associated
with $T_{1}$ and $T_{2}$ through the implicit solution of the equation
\begin{equation}
\tilde{\epsilon}(\rho _{1},\rho _{2},T,T)=\tilde{\epsilon}(\rho _{1},\rho
_{2},T_{1},T_{2}).  \label{averageT}
\end{equation}%
We denote by $\Theta _{\alpha }=T_{\alpha }-T$ the difference between
component and average temperatures which are non-equilibrium thermodynamical
variables. Near equilibrium, equation (\ref{averageT}) can be expanded to
the first order; then
\begin{equation}
\sum_{\alpha =1}^{2}c_{v}^{\alpha }\,\Theta _{\alpha }=0\qquad \mathrm{with}%
\qquad c_{v}^{\alpha }=\frac{\partial \widetilde{\epsilon }}{\partial
T_{\alpha }}(\rho _{1},\rho _{2},T,T).  \label{linearized}
\end{equation}%
Due to the fact that
\begin{equation*}
\rho \,d\varepsilon =\sum_{\alpha =1}^{2}\ \rho _{\alpha }\,T_{\alpha
}\,ds_{\alpha }+\frac{p_{\alpha }}{\rho _{\alpha }}\,d\rho _{\alpha },
\end{equation*}%
then
\begin{equation}
\rho \,c_{v}^{1}=T\sum_{\alpha =1}^{2}\rho _{\alpha }\,\frac{\partial {%
s_{\alpha }}}{\partial T_{1}}(\rho _{1},\rho _{2},T,T)\ \ \mathrm{and}\ \
\rho \,c_{v}^{2}=T\sum_{\alpha =1}^{2}\rho _{\alpha }\,\frac{\partial {%
s_{\alpha }}}{\partial T_{2}}(\rho _{1},\rho _{2},T,T).  \label{c}
\end{equation}%
The definition of the total entropy $s$ of the mixture is
\begin{equation}
\rho \,s=\sum_{\alpha =1}^{2}\rho _{\alpha }s_{\alpha }(\rho _{1},\rho
_{2},T_{1},T_{2}).  \label{average entropy3}
\end{equation}%
The first order expansion of equation (\ref{average entropy3}) yields
\begin{equation*}
\rho \,s=\sum_{\alpha =1}^{2}\rho _{\alpha }s_{\alpha }(\rho _{1},\rho
_{2},T,T)+\rho _{\alpha }s_{\alpha }\frac{\partial {s_{\alpha }}}{\partial
T_{1}}(\rho _{1},\rho _{2},T,T)\,\Theta _{1}+\rho _{\alpha }s_{\alpha }\frac{%
\partial {s_{\alpha }}}{\partial T_{2}}(\rho _{1},\rho _{2},T,T)\,\Theta
_{2}.
\end{equation*}%
Due to Relations (\ref{linearized}), (\ref{c})
\begin{equation*}
\rho \,s=\sum_{\alpha =1}^{2}\rho _{\alpha }s_{\alpha }(\rho _{1},\rho
_{2},T,T)
\end{equation*}%
and the specific entropy $s$ does not depend on $\Theta _{1}$ and $\Theta
_{2}$ but only on $\rho _{1},\rho _{2}$ and $T$. We denote by $\hat{%
\varepsilon}$ the internal specific energy as a function of $\rho ,c,T$:
\begin{equation*}
\hat{\varepsilon}(\rho ,c,T)=\tilde{\epsilon}(\rho _{1},\rho _{2},T,T),
\end{equation*}%
which satisfies the Gibbs equation
\begin{equation*}
Tds=d\hat{\varepsilon}-\frac{p_{o}}{\rho ^{2}}d\rho +\left( \mu _{2}-\mu
_{1}\right) dc
\end{equation*}%
where $p_{o}\left( \rho ,c,T\right) $ is the equilibrium pressure at
temperature $T$ and $\mu _{2}-\mu _{1}$, difference of component
chemical potentials,  is the chemical potential of the whole
mixture. By taking into account of equation (\ref{masses}), we get
\begin{equation*}
\rho \,\frac{d\hat{\varepsilon}}{dt}+p_{o}\ \mathrm{div}\,\mathbf{v}+\left(
\mu _{1}-\mu _{2}\right) \ \mathrm{div}\,\mathbf{j}-\rho \,T\,\frac{ds}{dt}%
=0.
\end{equation*}%
Moreover,
\begin{equation}
\rho \,\frac{ds}{dt}=\sum_{\alpha =1}^{2}\rho _{\alpha }\,\frac{d_{\alpha
}s_{\alpha }}{dt}+\mathrm{div}\,[(s_{2}-s_{1})\mathbf{j}\,].
\label{entropies}
\end{equation}

Equation (\ref{entropies}) yields the relation between the material
derivatives of entropy $s$ and entropies $s_{1}$ and $s_{2}$. By taking into
account of these results in equation (\ref{Total energy4}) and $\hat{\varepsilon}%
(\rho ,c,T)=\varepsilon (\rho _{1},\rho _{2},s_{1},s_{2})$, we obtain
\begin{equation*}
T\,\sum_{\alpha =1}^{2}\rho _{\alpha }\,\frac{d_{\alpha }s_{\alpha }}{dt}%
+\left( p-p_{o}\right) \ \mathrm{div}\,\mathbf{v}\,+\Big(
(h_{1}-h_{2})-\left( \mu _{1}-\mu _{2}\right) +T\,(s_{2}-s_{1})\Big) \,%
\mathrm{div}\,\mathbf{j}\,\
\end{equation*}%
\begin{equation}
+\big(\Theta _{1}\,\mathrm{{grad}\,\emph{s}_{1}-\Theta _{2}\,{grad}\,\emph{s}%
_{2}\big)^{T}\ \mathbf{j}}\ =\,0.  \label{Gibbs1}
\end{equation}

The differences of temperatures $\Theta _{1}\equiv T_{1}-T$ and $\Theta
_{2}\equiv T_{2}-T$ are small with respect to $T$ and $\mathbf{j}$%
\thinspace\ is a small diffusion term with respect to the mixture momentum $%
\rho \mathbf{v}$; consequently, in an approximation to the first order, the
term
\begin{equation*}
\big(\Theta _{1}\,\mathrm{{grad}\,\emph{s}_{1}-\Theta _{2}\,{grad}\,\emph{s}%
_{2}\big)^{\emph{T}}\ \mathbf{j}\ \,}
\end{equation*}%
is negligible. Let us consider
\begin{equation*}
K\equiv \Big((h_{1}-h_{2})-\left( \mu _{1}-\mu _{2}\right) +T\,(s_{2}-s_{1})%
\Big)\,\mathrm{div}\,\mathbf{j}\ ;
\end{equation*}%
we get
\begin{equation*}
K=\Big(\big(h_{1}-T_{1}s_{1}\big)-\big(h_{2}-T_{2}s_{2}\big)-\left( \mu
_{1}-\mu _{2}\right) +\Theta _{1}s_{1}+\Theta _{2}s_{2}\Big)\,\mathrm{div}\,%
\mathbf{j}\ .
\end{equation*}%
In an approximation to the first order, the term $\big(\Theta
_{1}s_{1}+\Theta _{2}s_{2}\big)\,\mathrm{div}\,\mathbf{j}\ $ is negligible.
\newline
Due to the fact that $\mu _{\alpha }(\rho _{1},\rho
_{2},T_{1},T_{2})=h_{\alpha }-T_{\alpha }s_{\alpha }$ is the chemical
potential of the component $\alpha $, when $\mathbf{j}$\thinspace\ is a
small diffusion velocity with respect to average velocity $\mathbf{v}$, the
term
\begin{equation*}
\displaystyle\Big( \mu _{1}(\rho _{1},\rho _{2},T_{1},T_{2})-\mu _{2}(\rho
_{1},\rho _{2},T_{1},T_{2})-\big( \mu _{1}\left( \rho _{1},\rho
_{2},T,T\right) -\mu _{2}\left( \rho _{1},\rho _{2},T,T\right) \big) \Big) \,%
\mathrm{div}\,\mathbf{j}
\end{equation*}%
is vanishing in an approximation to the first order.

Consequently, in an approximation to the first order, equation (\ref{Gibbs1}%
) reduces to
\begin{equation}
\sum_{\alpha =1}^{2}\rho _{\alpha }\,\frac{d_{\alpha }s_{\alpha }}{dt}=-%
\frac{1}{T}\left( p-p_{o}\right) \ \mathrm{div}\,\mathbf{v}.  \label{Planck}
\end{equation}%
The exchange of energy between components must obey the second law of
thermodynamics: the total entropy rate is an increasing function of time and
we consider the second law of thermodynamics in the form
\begin{equation}
\sum_{\alpha =1}^{2}\left( \frac{\partial \rho _{\alpha }s_{\alpha }}{%
\partial t}+\mathrm{{div}(\rho _{\alpha }s_{\alpha }\mathbf{v}_{\alpha })}%
\right) \,\geq 0  \label{CD1}
\end{equation}%
Due to relations (\ref{mass i}) the Clausius-Duhem inequality (\ref{CD1}) is
equivalent to
\begin{equation*}
\sum_{\alpha =1}^{2}\rho _{\alpha }\,\frac{d_{\alpha }s_{\alpha }}{dt}\,\geq
0\,.
\end{equation*}%
Relation (\ref{Planck}) implies that the second member must be positive.
Therefore, as usual in thermodynamics of irreversible processes, the entropy
inequality requires
\begin{equation}
\pi \equiv p-p_{o}=-\Lambda \ \mathrm{div}\,\mathbf{v}\,.
\label{other pressure}
\end{equation}%
This expression defines the Lagrange multiplier $\Lambda $ of
proportionality such that $\Lambda \geq 0$. The dynamical pressure $\pi $ is
the difference between the pressure in the process out of equilibrium with
different temperatures for the components and the pressure of the mixture
assumed in local thermodynamical equilibrium with the common average
temperature $T$. Let us notice that equations (\ref{total energy1},\ref%
{other pressure}) allow to obtain $Q_{\alpha }$ values. In fact,
\begin{equation*}
\rho _{1}T\,(T_{2}-T_{1})\frac{d_{1}\emph{s}_{1}}{dt}=\Lambda \,T_{2}\ (%
\mathrm{div}\,\mathbf{v)}^{2}\quad \text{\textrm{and}}\mathrm{\quad \rho _{2}%
\emph{T}\,(\emph{T}_{1}-\emph{T}_{2})\frac{d_{2}\emph{s}_{2}}{dt}=\Lambda \,%
\emph{T}_{1}\ (\mathrm{div}\,\mathbf{v)}^{2}}
\end{equation*}%
and the system of field equations is closed.

\section{Special case of mixture of perfect gases \protect\cite{RS}.}

The internal energy of the mixture is the sum of the internal energies of
the different gas components. We represent these energies as function of
density and temperature of components
\begin{equation*}
\widetilde{\varepsilon}(\rho_1,\rho_2,T_1,T_2) =\rho _{1}\,\widetilde{%
\varepsilon}_{1}(\rho _{1},T_{1})+\rho _{2}\, \widetilde{\varepsilon}%
_{2}(\rho _{2},T_{2}).
\end{equation*}%
Then,
\begin{equation*}
p=p_{1}(\rho _{1},T_{1})+p_{2}(\rho _{2},T_{2}),
\end{equation*}
where $p_1$ and $p_2$ are the pressures associated with $\widetilde{%
\varepsilon}_{1}$ and $\widetilde{\varepsilon}_{2}$. An expansion to the
first order in $T_1-T=\Theta_1$ and $T_2-T=\Theta_2$ yields
\begin{equation*}
p = p_{1}(\rho _{1},T)+p_{2}(\rho _{2},T)+\frac{\partial p_{1}}{\partial
T_{1}}(\rho _{1},T)\,\Theta_1+\frac{\partial p_{2}}{\partial T_{2}}(\rho
_{2},T)\,\Theta_2.
\end{equation*}%
From the definition of the average temperature $T$ we obtain as in \cite%
{Ruggeri3},
\begin{equation*}
\rho _{1}\,\widetilde{\varepsilon}_{1}(\rho _{1},T)+\rho _{2}\,\widetilde{%
\varepsilon}_{2}(\rho _{2},T)=\rho _{1}\,\widetilde{\varepsilon}_{1}(\rho
_{1},T_{1})+\rho _{2}\,\widetilde{\varepsilon}_{2}(\rho _{2},T_{2}).
\end{equation*}%
Consequently, from equation (\ref{other pressure}), we get
\begin{equation*}
\pi =\frac{\partial p_{1}}{\partial T_{1}}(\rho _{1},T)\,\Theta_1+\frac{%
\partial p_{2}}{\partial T_{2}}(\rho _{2},T)\,\Theta_2.
\end{equation*}%
Let $T_{1}=T+\beta \,\Theta ,\ T_{2}=T+(1+\beta )\,\Theta ,$ where $\Theta
=T_{2}-T_{1}$, an expansion of equation (17) to the first order yields the
value of $\beta $
\begin{equation*}
C_{v}^{(1)}(\rho _{1},T)\,\beta \,\Theta +C_{v}^{(2)}(\rho _{2},T)\,(1+\beta
)\,\Theta =0,\ \ \mathrm{with}\ \ C_{v}^{(\alpha )}=\frac{\partial
\widetilde{\varepsilon}_{\alpha}}{\partial T_\alpha}(\rho _{\alpha },\emph{T}%
).
\end{equation*}%
Consequently, when $p_{\alpha }=k_{\alpha }\rho _{\alpha }T_{\alpha }$, we
obtain
\begin{equation*}
\pi =\frac{\rho _{1}\rho _{2}}{\rho _{1}C_{v}^{(1)}+\rho _{2}\,C_{v}^{(2)}}%
\,(k_{2}\,C_{v}^{(1)}-k_{1}C_{v}^{(2)})\,\Theta
\end{equation*}%
In accordance with results obtained by Ruggeri \& Simi\'{c} \cite{Ruggeri2}
and Gouin \& Ruggeri \cite{Gouin-Ruggeri}, to verify the Clausius-Duhem
inequality (\ref{CD1}), $\Theta $ must be in the form
\begin{equation*}
\Theta =L_{T}(\gamma _{1}-\gamma _{2})\,div\,\mathbf{v}\quad \mathrm{with}
\quad L_{T}=M\,\frac{\rho _{1}C_{v}^{(1)}}{\rho _{2}C_{v}^{(2)}}\,(\rho
_{1}C_{v}^{(1)}+\rho _{2}C_{v}^{(2)})\ \ \mathrm{and}\ \ M\geq 0
\end{equation*}%
where $\gamma _{\alpha }$ is the ratio of specific heats of component $%
\alpha $.

\section{Conclusion.}

The method by Hamilton can be easily extended to multi-component mixtures
with multi-temperatures. We obtain the equations of component motions and
the equation of the total mixture energy. The entropy is not conserved and
the second law of thermodynamics reveals the existence of a new dynamical
pressure term. As diffusion is a property of fluid mixtures with different
component velocities, the dynamical pressure term is a property of fluid
mixtures with different component temperatures. The dynamical pressure can
be measured with the change of volume. In the special case of mixture of
gases, the dynamical pressure term comes from the fact the gases are
molecularly different. \newline
The Hamilton principle points out that the dynamical pressure can be
obtained by neglecting viscosity, friction or external heat fluxes. This is
a main property of mixtures with multi-temperatures and this fact may have
some applications in plasma of gases and in the evolution of the early
universe \cite{Weinberg}. \newline
In Appendix C, we highlight that constitutive equations for diffusion,
viscosity and heat flux for mixtures without chemical reaction are
consequence of dissipative terms whereas the dynamical pressure term can
exist with different component temperatures even if the bulk viscosity is
null.\newline
The results are in complete accordance with the ones by Ruggeri \& Simic
\cite{Ruggeri2} and Gouin \& Ruggeri \cite{GR}. This is an important
verification of the fact that the Hamilton principle can be extended to
nonconservative mixture motions when components have different temperature.
A difference with classical thermodynamics methods  is that the volume
internal energy is not necessary the sum of the volume internal energies of
the components. In this paper, the volume internal energy is a nonseparate
function of densities and entropies (or temperatures) and is consequently
more general than in \cite{GR} and \cite{Ruggeri2}. \newline

\bigskip

\centerline{{\bf
Appendix A}.} \bigskip \noindent The definition of Lagrangian coordinates $\,%
{\mathbf{X}}_{\alpha }\,$ implies\ \ $\displaystyle\frac{\partial {\mathbf{X}%
}_{\alpha }}{\partial t}\,+\,\frac{\partial {\mathbf{X}}_{\alpha }}{\partial
{\mathbf{x}}}\ {\mathbf{v}}_{\alpha }\,=\,0$. By taking the derivative with
respect to $\,\varkappa _{\alpha }$, we obtain the following equation for
virtual displacements (equation (\ref{displacement1})) associated with the
first virtual motion family
\begin{equation*}
\frac{\partial {\delta _{\alpha }}{\mathbf{X}}_{\alpha }}{\partial t}\ +\
\frac{\partial {\delta _{\alpha }}{\mathbf{X}}_{\alpha }}{\partial {\mathbf{x%
}}}\ {\mathbf{v}}_{\alpha }\,+\,\frac{\partial {{\mathbf{X}}_{\alpha }}}{%
\partial {\mathbf{x}}}\ {\delta _{\alpha }}{\mathbf{v}}_{\alpha }\ =\ 0.
\end{equation*}%
Then, we get
\begin{equation*}
\ {\delta _{\alpha }}{\mathbf{v}}_{\alpha }\,({\mathbf{x}},t)\,=\,-\
F_{\alpha }\,\frac{d_{\alpha }}{dt}\ ({\delta _{\alpha }}{\mathbf{X}}%
_{\alpha }).
\end{equation*}%
Equation (\ref{mass0}) yields
\begin{equation}
{\delta _{\alpha }}\rho _{\alpha }\,({\mathbf{x}},t)\,\det \,F_{\alpha }\,({%
\mathbf{x}},t)\,+\,\rho _{\alpha }\,{\delta _{\alpha }}\,(\det \,F_{\alpha
})\,=\,\frac{\partial \rho _{\alpha 0}}{\partial {\mathbf{X}}_{\alpha }}\,{%
\delta _{\alpha }}{\mathbf{X}}_{\alpha }.  \label{(transform)}
\end{equation}%
By using the Euler-Jacobi identity
\begin{equation*}
\displaystyle{\delta _{\alpha }}(\det \ F_{\alpha })\,=\,\det \,F_{\alpha
}\,({\mathbf{x}},t)\ \ tr\ \bigg(F_{\alpha }^{-1}\ {\delta _{\alpha }}%
F_{\alpha }\ \bigg)
\end{equation*}%
with
\begin{equation*}
\displaystyle{\delta _{\alpha }}F_{\alpha }\ ({\mathbf{x}},t)\,=\,-\
F_{\alpha }\ ({\mathbf{x}},t)\ {\delta _{\alpha }}F_{a}^{-1}\ ({\mathbf{x}}%
,t)\ F_{\alpha }\ ({\mathbf{x}},t)
\end{equation*}%
and
\begin{equation*}
{\delta _{\alpha }}F_{a}^{-1}\,({\mathbf{x}},t)\,=\,\frac{\partial {\delta
_{\alpha }}{\mathbf{X}}_{\alpha }}{\partial {\mathbf{x}}},
\end{equation*}%
we deduce
\begin{equation*}
{\delta _{\alpha }}(\det \ F_{\alpha }\ )=-\det \ F_{\alpha }\ \ tr\ \bigg({%
\delta _{\alpha }}F_a^{-1}\ F_{\alpha }\bigg)\ =-\det \ F_{\alpha }\ tr\bigg(\,%
\frac{\partial {\ {\delta _{\alpha }}{\mathbf{X}_{\alpha }}}}{\partial {%
\mathbf{X}}_{\alpha }}\bigg),
\end{equation*}%
or,
\begin{equation}
{\delta _{\alpha }}(\det F_{\alpha })\ =\ -\det \,F_{\alpha }\ \ div_{\alpha
}({\delta _{\alpha }}{\mathbf{X}}_{\alpha }).  \label{(legendre2)}
\end{equation}%
By substituting equation (\ref{(legendre2)}) into equation (\ref{(transform)}%
), we obtain
\begin{equation*}
{\delta _{\alpha }}\rho _{\alpha }({\mathbf{x}},t)\ =\ \rho _{\alpha }\
div_{\alpha }({\delta _{\alpha }}{\mathbf{X}}_{\alpha })\,+\,\,\frac{\rho
_{\alpha }}{\rho _{\alpha 0}}\ \frac{\partial \rho _{\alpha 0}}{\partial {%
\mathbf{X}}_{\alpha }}\ {\delta _{\alpha }}{\mathbf{X}}_{\alpha }=\frac{%
div_{\alpha }(\rho _{\alpha 0}\,{\delta _{\alpha }}{\mathbf{X}}_{\alpha })}{%
detF_{\alpha }},
\end{equation*}%
\begin{equation*}
{\delta _{\alpha }}s_{\alpha }\,({\mathbf{x}},t)\ =\frac{\partial s_{\alpha
0}}{\partial {\mathbf{X}}_{\alpha }}\ {\delta _{\alpha }}{\mathbf{X}}%
_{\alpha }.
\end{equation*}

{\small \bigskip }

{\small \centerline{{\bf Appendix B}.} }

{\small \bigskip }

{\small \noindent The proof of the Gibbs identity is obtained by summing the
following algebraic identities $\ a - e , \ $ }

{\small \noindent For the external potentials $\ \Omega_\alpha $, }

{\small \noindent \textbf{a.}
\begin{equation*}
\displaystyle \ {\frac{\partial\, \rho_\alpha\Omega_\alpha }{\partial t}} +\
div\ (\rho_\alpha\Omega_\alpha{\mathbf{v}}_\alpha) \ -\ \rho_\alpha\ {\frac{%
\partial \Omega_\alpha}{\partial {\mathbf{x}}}}\ {\mathbf{v}}_\alpha \ -\
B_\alpha\Omega_\alpha \ -\ \rho_\alpha\ {\frac{\partial \Omega_\alpha}{%
\partial t}}\ \equiv\ 0
\end{equation*}
\noindent For the velocity fields $\ {\mathbf{v}}_\alpha,$ }

{\small \noindent \textbf{b.}
\begin{equation*}
\displaystyle\,{\frac{\partial }{\partial t}}\ \left( {\frac{1}{2}}\ \rho
_{\alpha }\ {\mathbf{v}}_{\alpha }^{2}\right) \ +\ div\ \left( \rho _{\alpha
}\ {\mathbf{v}}_{\alpha }\, \left({\mathbf{v}}_{\alpha }^{2}-{\frac{1}{2}}\ {%
\mathbf{v}}_{\alpha }^{2}\right)\right) \
\end{equation*}%
\begin{equation*}
\displaystyle-\ B_{\alpha }\ \left( {\mathbf{v}}_{\alpha }^{2}\ -\ {\frac{1}{%
2}}\ {\mathbf{v}}_{\alpha }^{2}\right) -\left( \rho _{\alpha }\ {\frac{%
d_{\alpha }{\mathbf{v}}_{\alpha }^{T}}{dt}}\ +\ \rho _{\alpha }\ {\mathbf{v}}%
_{\alpha }^{T}\ {\frac{\partial {\mathbf{v}}_{\alpha }}{\partial {\mathbf{x}}%
}}-\rho _{\alpha }\ {\frac{\partial }{\partial {\mathbf{x}}}}\, \left({\frac{%
1}{2}}\ {\mathbf{v}}_{\alpha }^{2}\right)\right) \ {\mathbf{v}}_{\alpha }\
\equiv \ 0
\end{equation*}%
Let us introduce $\displaystyle\ {\mathbf{i}}^{T}\ =\ -\ {\frac{\partial
\eta }{\partial {\mathbf{u}}}}$ . Then
\begin{equation*}
\displaystyle {\frac{\partial }{\partial t}}\,\bigg(\eta -{\frac{\partial
\eta }{\partial {\mathbf{u}}}}\ {\mathbf{u}}\bigg)\ =\ {\frac{\partial \ {%
\mathbf{i}}^{T}}{\partial t}}\ {\mathbf{u}}\ +\ \sum_{\alpha =1}^{2}\ \bigg(%
\ {\frac{\partial \eta }{\partial \rho _{\alpha }}}\ {\frac{\partial \rho
_{\alpha }}{\partial t}}\ +\ \rho _{\alpha }\ T_{\alpha }\ {\frac{\partial
s_{\alpha }}{\partial t}}\ \bigg)
\end{equation*}%
\noindent and the three following identities $\ c-e\ $ prove the formula}

{\small \noindent \textbf{c.}
\begin{equation*}
\displaystyle {\frac{\partial \eta}{\partial \rho_\alpha}}\ {\frac{\partial
\rho_\alpha}{\partial t}} + div \left( {\frac{\partial \eta}{\partial
\rho_\alpha}}\ \rho_\alpha\ {\mathbf{v}}_\alpha\right) - \rho_\alpha\ {\frac{%
\partial }{\partial {\mathbf{x}}}}\ \bigg(\ {\frac{\partial \eta}{\partial
\rho_\alpha}}\bigg)\ {\mathbf{v}}_\alpha - {\frac{\partial \eta}{\partial
\rho_\alpha}}\ \bigg({\frac{\partial \rho_\alpha}{\partial t}} \ +\ div\
(\rho_\alpha\ {\mathbf{v}}_\alpha)\bigg)\ \equiv \ 0,
\end{equation*}
}

{\small \noindent \textbf{d.}
\begin{equation*}
\displaystyle \rho_\alpha\ T_\alpha\ {\frac{\partial s_\alpha}{\partial t}}
\ +\ \rho_\alpha\ T_\alpha\ {\frac{\partial s_\alpha}{\partial {\mathbf{x}}}}%
\ {\mathbf{v}}_\alpha \ -\ \rho_\alpha\ T_\alpha\ {\frac{d_\alpha s_\alpha}{%
dt}}\ \equiv \ 0,
\end{equation*}
}

{\small \noindent \textbf{e.}
\begin{equation*}
{\frac{\partial {\mathbf{i}}^{T}}{\partial t}}\ {\ \eta } \ +\sum_{\alpha
=1}^{2}\,div\left( (-1)^{\alpha }\left( {\frac{{\mathbf{i}}^{T}}{\rho
_{\alpha }}}\,{\mathbf{v}}_{\alpha }\right) \rho _{\alpha }{\mathbf{v}}%
_{\alpha }\right) -\left( \rho _{\alpha }{\frac{d_{\alpha }}{dt}}\left(
(-1)^{\alpha }\,{\frac{{\mathbf{i}}^{T}}{\rho _{\alpha }}}\ \right) +\rho
_{\alpha }(-1)^{\alpha }{\frac{{\mathbf{i}}^{T}}{\rho _{\alpha }}}\,{\frac{%
\partial {\mathbf{v}}_{\alpha }}{\partial {\mathbf{x}}}}\right) \,{\mathbf{v}%
}_{\alpha }
\end{equation*}%
\begin{equation*}
\displaystyle-\ (-1)^{\alpha }\ \left( {\frac{{\mathbf{i}}^{T}}{\rho
_{\alpha }}}\ {\mathbf{v}}_{\alpha }\right) \,\left( \ {\frac{\partial \rho
_{\alpha }}{\partial t}}+div\,(\rho _{\alpha }\ {\mathbf{v}}_{\alpha
})\right) \ \equiv \ 0 .
\end{equation*}
}

{\small \bigskip  }

{\small \centerline{{\bf Appendix C}.}  }

{\small \bigskip  }

{\small We consider a more general case of a mixture when the
Hamilton principle cannot be applied. This case consists of a weak
dissipative process with diffusion, viscosity and heat transfers.
The balance of masses, momenta and energy are simply expressed by
adding dissipative terms to the expressions obtained in section 4
\begin{equation}
B_{\alpha }=0,\quad {\mathbf{M}}_{\alpha }^{d}=0\ \ (\alpha =1,2)\quad \text{%
and}\quad E^{d}=0,  \label{nonconservative equations}
\end{equation}%
such that
\begin{eqnarray*}
&&\displaystyle B_{\alpha }\ =\frac{\partial \rho _{\alpha }}{\partial t}\
+\ {\mathrm{div}(\rho _{\alpha }{\mathbf{v}}_{\alpha }),} \\
&&\mathbf{M}_{\alpha }^{d}=\mathbf{M}_{\alpha }-(\mathrm{div\ }\mathbf{%
\sigma }_{\alpha }^{d})^{T}-\ \mathbf{m}_{\alpha },\ \ \ \  \\
&&E^{d}=E+\sum_{\alpha =1}^{2}\mathrm{div\ }\mathbf{q}_{\alpha }-\mathbf{%
\sigma }_{\alpha }^{d}{\mathbf{v}}_{\alpha }.
\end{eqnarray*}%
On the right hand side, $\mathbf{q}_{\alpha }$ is the heat flux vector, $%
\mathbf{m}_{\alpha }$ is the momentum production and $\mathbf{\sigma }%
_{\alpha }^{d}$ is the viscous part of the stress tensor of constituent $%
\alpha $. Due to the total conservation of momentum of the mixture, $%
\sum_{\alpha =1}^{2}\ \mathbf{m}_{\alpha }=0\ \ $\cite{ET}.\newline
The \textit{dynamics} \textit{Gibbs identity} (\ref{Identity of Gibbs}) can
be transformed as
\begin{equation*}
S^{d}+\sum_{\alpha =1}^{2}\mathbf{M}_{\alpha }^{d\,T}\,\mathbf{v}_{\alpha
}-\left( \frac{1}{2}\,\mathbf{v}_{\alpha }^{2}-h_{\alpha }+T_{\alpha
}s_{\alpha }\right) \,B_{\alpha }\equiv \ E^{d},
\end{equation*}%
with
\begin{equation*}
S^{d}=S+\mathrm{div\ }\mathbf{q}+\sum_{\alpha =1}^{2}\mathbf{m}_{\alpha
}^{T}\,\mathbf{v}_{\alpha }-\mathrm{tr}\ (\mathbf{\sigma }_{\alpha }^{d}\
\mathbf{D}_{\alpha }),
\end{equation*}%
where $\displaystyle\mathbf{D}_{\alpha }=\dfrac{1}{2}\left( \frac{\partial
\mathbf{v}_{\alpha }}{\partial \mathbf{x}}+\left( \frac{\partial \mathbf{v}%
_{\alpha }}{\partial \mathbf{x}}\right) ^{T}\right) \ $and $\displaystyle%
\mathbf{q}=\sum_{\alpha =1}^{2}\mathbf{q}_{\alpha }$.\newline
The second law of thermodynamics is expressed in the form
\begin{equation}
\sum_{\alpha =1}^{2}\rho _{\alpha }\frac{d_{\alpha }s_{\alpha }}{dt}+\mathrm{%
div\ }\frac{\mathbf{q}_{\alpha }}{T_{\alpha }}\geq 0.  \label{CD3}
\end{equation}%
In the second order approximation, with small external heat fluxes $\mathbf{q%
}_{\alpha }$ and small difference of temperature $T_{1}-T_{2}$, equation (%
\ref{CD3}) is equivalent to
\begin{equation*}
\sum_{\alpha =1}^{2}\rho _{\alpha }\frac{d_{\alpha }s_{\alpha }}{dt}+\mathrm{%
div\ }\frac{\mathbf{q}}{T}\geq 0.
\end{equation*}%
If we write $\mathbf{m}=\mathbf{m}_{1}=-\mathbf{m}_{2}$, we obtain by
calculations similar to section\ 4
\begin{equation*}
T\,\left[ \sum_{\alpha =1}^{2}\rho _{\alpha }\,\frac{d_{\alpha }s_{\alpha }}{%
dt}+\ \mathrm{div\ }\frac{\mathbf{q}}{T}\right] +\Sigma =0,
\end{equation*}%
where
\begin{equation*}
\Sigma =\left( p-p_o\right) \
\mathrm{div}\,\mathbf{v}\,+\frac{\mathbf{q}}{T}\,\mathrm{{grad}\,T-\mathbf{m}%
^{T}\mathbf{u}-\sum_{\alpha =1}^{2}\mathrm{tr}\ (\mathbf{\sigma }_{\alpha
}^{d}\ \mathbf{D}_{\alpha })\leq 0}
\end{equation*}%
is the entropy production. \newline
Classical methods of thermodynamics of irreversible process (\emph{TIP})
yield equation (\ref{other pressure}) for the dynamical pressure term
together with Fourier and Navier-Stokes laws \cite{Muller}.\newline
Term $\mathbf{m}^{T}\mathbf{u}$ yields the coefficient$_{{}}$ $\chi $ of
proportionality such that $\chi \geq 0$ and
\begin{equation*}
\mathbf{m}=-\chi \,\mathbf{u}\equiv \chi \,(\mathbf{v}_{1}-\mathbf{v}_{2}).
\end{equation*}%
For slow isothermal motions, the difference between the components of
equations (\ref{nonconservative equations})$_{2}$ yields in an approximation
to the first order
\begin{equation*}
\frac{\mathbf{M}_{1}}{\rho _{1}}-\frac{\mathbf{M}_{2}}{\rho _{2}}=\mathrm{{%
grad}(\mu _{1}-\mu _{2}).}
\end{equation*}%
Here $\displaystyle\mu _{\alpha }=\frac{\partial \eta }{\partial \rho
_{\alpha }}-T_{\alpha }s_{\alpha }$ denotes the chemical potential of
component $\alpha $ at temperature $T$.\newline
By neglecting the viscous terms, we obtain
\begin{equation*}
\mathrm{{grad}\,\mu =\frac{\rho }{\rho _{1}\rho _{2}}\ \mathbf{m}\quad \
(\mu =\mu _{1}-\mu _{2})\quad \text{ \ or }\quad {grad}\,\mu =-\kappa \,%
\mathbf{u}\quad \text{ with}\quad \kappa =\frac{\rho }{\rho _{1}\rho _{2}}%
\,\chi ,}
\end{equation*}%
which is an expression of the Fick law.\newline
Therefore, in this formulation, our results coincide with the ones obtained
by arguments of classical thermodynamics \cite{GR}. \newline
}

{\small \bigskip }

{\small \text{\textbf{Acknowledgements}} }

{\small {This paper was developed during a stay of\, T.R. as visiting
professor at the University of Aix-Marseille and was supported in part
(T.R.) by fondi MIUR Progetto di interesse Nazionale \emph{Problemi
Matematici Non Lineari di Propagazione e Stabilit\`{a} nei Modelli del
Continuo} Coordinatore T. Ruggeri, by the GNFM-INdAM. }  }


\begin{thebibliography}{99}
\bibitem{GR} {\small H. Gouin, T. Ruggeri, Identification of an average
temperature and a dynamical pressure in multi-temperature mixture of fluids,
\textit{Physical Review E} \textbf{78} (2008) 016303. }

\bibitem{Ishii} {\small M. Ishii, \textit{Thermo-fluid dynamic theory of
two-phase flows}, Paris, Eyrolles, 1975.  }

\bibitem{Nigmatulin} {\small R.I. Nigmatulin, \textit{Dynamics of multiphase
media}, Hemisphere Publ. Corp., New York, 1991.  }

\bibitem{Drew} {\small D.A. Drew, Mathematical modeling of two-phase flow,
\textit{Annual Rev. Fluid Mech.} \textbf{15} (1983) 261-291.  }

\bibitem{Win} {\small A. Biesheuvel, L. van Wijngaarden, Two-phase flow
equations for a dilute dispersion of gas bubbles in liquid, \textit{J. Fluid
Mech.} \textbf{148} (1984) 301-318.  }

\bibitem{Bowen} {\small R.M. Bowen, \textit{Theory of mixtures}, in
Continuum Physics, Vol III, A.C. Eringen Ed., Academic Press, London, 1976,
pp. 1-127.  }

\bibitem{Landau} {\small L.D. Landau, E. Lifshits, \textit{Fluid mechanics},
Pergamon Press, London, 1989.  }

\bibitem{Putterman} {\small S. Putterman, \textit{Superfluid hydrodynamics},
Elsevier, New York, 1974.  }

\bibitem{Khalatnikov} {\small I.M. Khalatnikov, \textit{Theory of
superfluidity}, Nauka, Moscow, 1971.  }

\bibitem{Truesdell} {\small C. Truesdell, \textit{Rational thermodynamics},
Series in Modern Applied Mathematics, MacGraw-Hill, New York, 1969. }

\bibitem{Muller} {\small I. M\"{u}ller, \textit{Thermodynamics}, Interaction
of Mechanics and Mathematics Series, Pitman, London, 1985.  }

\bibitem{ET} {\small I. M\"{u}ller, T. Ruggeri, \textit{Rational extended
thermodynamics}, Springer, New York, 1998.  }

\bibitem{RS} {\small T. Ruggeri, S. Simi\'{c}, On the hyperbolic system of a
mixture of Eulerian fluids: a comparison between single and
multi-temperature models, \textit{Math. Meth. Appl. Sci.} \textbf{30 }(2007)
827-849.  }

\bibitem{Lin} {\small C.C. Lin, A new variational principle for isoenergetic
flows, \textit{Q. J. Appl. Math.} \textbf{9} (1952) 421-423.  }

\bibitem{Serrin} {\small J. Serrin, \textit{Mathematical principles of
classical fluid mechanics}, Encyclopedia of Physics VIII/1, S. Fl\"{u}gge
Ed., Springer, Berlin, 1960, pp. 125-263.  }

\bibitem{gouin0} {\small {\ H. Gouin,} {Thermodynamic form of the equation
of motion for perfect fluids of grade n,} \textit{Comptes Rendus Acad. Sci.
Paris} \textbf{305}, II (1987) 833-838.  }

\bibitem{gouin1} {\small {\ H. Gouin, J.F. Debieve,} Variational principle
involving the stress tensor in elastodynamics, \textit{Int. J. Engng. Sci.}
\textbf{24} (1986) 1057-1069.  }

\bibitem{Bedford} {\small A. Bedford, D.S. Drumheller, Recent advances.
Theories of immiscible and structured mixtures, \textit{Int. J. Engng. Sci.}
\textbf{21} (1983) 863-960.  }

\bibitem{Berdichevsky} {\small V.L. Berdichevsky, \textit{Variational
principles of continuum mechanics}, Nauka, Moscow, 1983.  }

\bibitem{Geurst2} {\small J. A. Geurst, Variational principles and two-fluid
hydrodynamics of bubbly liquid/gas mixtures, \textit{Physica A} \textbf{135}
(1986) 455-486.  }

\bibitem{gouin2} {\small {\ H. Gouin,} Variational theory of mixtures in
continuum mechanics, \textit{Eur. J. Mech. B/Fluids} \textbf{9} (1990)
469-491.  }

\bibitem{Gavrilyuk 1} {\small S. Gavrilyuk, H. Gouin, Yu. Perepechko, A
variational principle for two-fluid models, \textit{Comptes Rendus Acad.
Sci. Paris} \textbf{324}, II (1997) 483-490.  }

\bibitem{Gouin-Ruggeri} {\small {\ H. Gouin, T. Ruggeri,} Mixture of fluids
involving entropy gradients and acceleration waves in interfacial layers,
\textit{Eur. J. Mech. B/Fluids} \textbf{24} (2005) 596-613.  }

\bibitem{gouin3} {\small H. Gouin, S. Gavrilyuk, Hamilton's principle and
Rankine-Hugoniot conditions for general motions of mixtures, \textit{%
Meccanica} \textbf{34} (1999) 39-47.  }

\bibitem{Gavrilyuk 2} {\small S. Gavrilyuk, H. Gouin, Yu. Perepechko,
Hyperbolic models of homogeneous two-fluid mixtures, \textit{Meccanica}
\textbf{33} (1998) 161-175.  }

\bibitem{lhuillier} {\small D. Lhuillier, From molecular mixtures to
suspensions of particles,\textit{\ J. Physique II} \textbf{5} (1995) 19-36.
}

\bibitem{Groot} {\small S.R. de Groot, W.A. van Leeuwen, Ch.G. van Weert,
\textit{Relativistic kinetic theory}. North-Holland, Amsterdam, 1980.  }

\bibitem{Stewart} {\small H.B. Stewart, B. Wendroff, Two-phase flow: models
and methods, \textit{J. Comput. Phys.} \textbf{56} (1984) 363-409.  }

\bibitem{Ruggeri3} {\small T. Ruggeri, S. Simi\'{c}, \textit{Mixture of
gases with multi-temperature: Identification of a macroscopic average
temperature}, in: Proceedings of Workshop Mathematical Physics Models and
Engineering Sciences, 455-462, Liguori Editore, Napoli (2008).  }

\bibitem{Ruggeri2} {\small T. Ruggeri, S. Simi\'{c}, \textit{Mixture of
gases with multi-temperature: Maxwellian iteration}, in: Asymptotic Methods
in Non Linear Wave Phenomena, T. Ruggeri and M. Sammartino Eds., 186-194,
World Scientific, Singapore (2007).  }

\bibitem{Gavrilyuk 3} {\small {\ S. Gavrilyuk, H. Gouin, A new form of
governing equations of fluids arising from Hamilton's principle}, \textit{%
Int. J. Engng. Sci.} \textbf{37} (1999) 1495-1520.  }

\bibitem{Weinberg} {\small S. Weinberg, Entropy generation and the survival
of protogalaxies in an expanding universe, \textit{Astrophys. J.} \textbf{168%
} (1971) 175-194.  }
\end{thebibliography}
\end{document}